# Electronic and Magnetic Properties of Topological Semimetal Candidate NdSbTe

Krishna Pandey[1], Rabindra Basnet[2*], Aaron Wegner[2], Gokul Acharya[2], Md Rafique Un Nabi[2], Jiangwei Liu[3], Jian Wang[4], Yukiko Takahashi[5], Bo Da[6], Jin Hu[1,2‡]

[1]Materials Science and Engineering Program, Institute for Nanoscience and Engineering, University of Arkansas, Fayetteville, AR 72701, USA

[2]Department of Physics, University of Arkansas, Fayetteville, AR 72701, USA

[3]Research Center for Functional Materials, National Institute for Materials Science, Tsukuba, Ibaraki 305-0047, Japan

[4]Institute for Materials Research and Center for Spintronics Research Network, Tohoku University, Sendai 980-8577, Japan

[5]Research Center for Magnetic and Spintronic Materials, National Institute for Materials Science, Sengen 1-2-1, Tsukuba, Ibaraki 305-0047, Japan

[6]Research and Services Division of Materials Data and Integrated System, National Institute for Materials Science, Tsukuba, Ibaraki 305-0044, Japan

## Abstract

ZrSiS-type materials represent a large material family with unusual coexistence of topological nonsymmorphic Dirac fermions and nodal-line fermions. As a special group of ZrSiS-family, $Ln$SbTe ($Ln$ = Lanthanide rare earth) compounds provide a unique opportunity to explore new quantum phases due to the intrinsic magnetism induced by $Ln$. Here we report the single crystal growth and characterization of NdSbTe, a previously unexplored $Ln$SbTe compound. NdSbTe has an antiferromagnetic ground state with field-driven metamagnetic transitions similar to other known $Ln$SbTe, but exhibits distinct enhanced electronic correlations characterized by large a Sommerfeld coefficient of 115 mJ/mol K$^2$, which is the highest among the known *Ln*SbTe compounds. Furthermore, our transport studies have revealed the coupling with magnetism and signatures of Kondo localization. All these findings establish NdSbTe as a new platform for observing novel phenomena arising from the interplay between magnetism, topology, and electron correlations.

Email: [*]rbasnet@uark.edu; [‡]jinhu@uark.edu

Topological semimetals (TSMs) have gained increased attention due to recent breakthroughs [1-3]. These materials host relativistic fermions with low-energy electronic excitations that resemble the Dirac and Weyl fermions of the Standard Model, exhibit various technologically useful properties such as large magnetoresistance and ultrahigh carrier mobility [4], and possess exotic quantum phenomena such as a chiral anomaly [5-9] and unusual Fermi arcs [10-14]. To date, many TSMs have been predicted and experimentally demonstrated [1-3,15-19]. Among these, there has recently been rapidly growing interest in the ZrSiS-family of topological materials. ZrSiS, the representative material of this family, crystallizes in the tetragonal PbFCl-type structure (space group $P4/nmm$) made up of two-dimensional (2D) square nets of Si sandwiched by Zr-S layers that host a 2D gapless nonsymmorphic Dirac state [20] predicted by Young and Kane [21]. Additionally, linear band crossings close to the Fermi level that form a Dirac nodal-line have been observed in ZrSiS, which is slightly gapped by weak spin-orbit coupling [20,22,23]. Other materials in this family known to host topological fermions include Zr(Si/Ge)(S/Se/Te) [24-27], Hf(Si/Ge)(S/Se/Te) [28-31], and ZrSnTe [32,33]. This abundance of isomorphic topological materials provides a great opportunity to fine tune the topological states by varying spin-orbital coupling, electronic dimensionality, and lattice constant through element substitutions. These materials are also of particular interest for thin film applications, as some of them can be exfoliated due to their layered crystal structure with weak interlayer binding [27,34], which can lead to other exotic phases such as quantum spin Hall insulator in the monolayer form [22].

While the square net planes formed by Group-IV elements (Si, Ge, and Sn) play an essential role in generating the topologically non-trivial bands in the above mentioned materials, the Sb (Group-V) network can also supports topological fermions in the related compound of this family, *Ln*SbTe (*Ln* = Lanthanide rare earth). Thus far, only a few *Ln*SbTe compounds have been studied [22,35-42]. The orthorhombically distorted non-magnetic LaSbTe has been suggested to be a topological insulator [22,36], and a topological nodal-line in GdSbTe has been observed by angle-resolved photoemission spectroscopy (ARPES) [38]. Furthermore, the rich magnetic phases of CeSbTe, which originate from the 4*f*-magnetism of Ce, have led to predictions of Dirac and time-reversal breaking Weyl states tunable by temperature and magnetic field [37]. Given the rich magnetic properties [35,39-41] and large material pool available by varying *Ln*, this less-explored *LnSbTe*-family provides a rare platform for investigating the interplay between magnetism and electronic band topology.

With this motivation, we synthesized single crystals of the previously unexplored NdSbTe and characterized their magnetic, calorimetric, and electronic transport properties. We found this material has an antiferromagnetic (AFM) ground state with field-induced metamagnetic transitions that couple to the electronic transport. The Kondo-like transport feature, together with the significantly enhanced electronic correlation, implies that NdSbTe is a model system exhibiting interplay between magnetism, topology, and electronic correlations, which may further give rise to new quantum phenomena such as a Kondo-Weyl state and correlated topological phases.

The NdSbTe single crystals used in this work were synthesized by a chemical vapor transport method with $I_2$ as the transport agent. First, a polycrystalline precursor of NdSbTe was prepared by heating the stoichiometric mixture of Nd, Sb and Te powder at 750 °C for 2 days. The precursor was then used as a source for chemical vapor transport with a temperature gradient from 1000 °C to 850 °C for 2 weeks. Millimeter-size single crystals with metal luster were obtained (Fig. 1, inset). The composition and structure of the synthesized crystals were examined by energy-dispersive x-ray spectroscopy and x-ray diffraction, respectively. Magnetization, specific heat, and electronic transport were measured using a physical property measurement system (PPMS, Quantum Design).

The single crystal x-ray diffraction pattern for NdSbTe is shown in Fig. 1, which reveals excellent crystallinity as demonstrated by the sharp (00$L$) peaks. The extracted *c*-axis lattice parameter of 9.36 Å is in between that of CeSbTe [37,39] and GdSbTe [38,40,41], which is consistent with the ionic radius of these rare earth elements ($Ce^{3+}$ > $Nd^{3+}$ > $Gd^{3+}$) and implies a structural similarity of these compounds. As ZrSiS-type materials [43], the crystal structures of *Ln*SbTe compounds consist of stacking Te-*Ln*-Sb-*Ln*-Te slabs along the *c*-axis, with Sb plane sandwiched by *Ln*-Te layers (Fig. 1, inset). The Sb planar layer is similar to the Si square net in ZrSiS that hosts 2D/quasi-2D relativistic fermions [20,22,23]. On the other hand, the Sb layer can be partially substituted by Te, leading to non-stoichiometric composition of $Ln\text{Sb}_{1-x}\text{Te}_{1+x}$, which has been widely seen for the currently known magnetic *Ln*SbTe compounds such as CeSbTe [39] and GdSbTe [40]. The composition non-stoichiometry can give rise to an orthorhombic distortion in *Ln*SbTe [40], though the distortions do not appear to affect the presence of topological fermions [36,38,40]. In our single crystal study, non-stoichiometric $\text{NdSb}_{1-x}\text{Te}_{1+x}$ with $x$ ranging from 0.2 ~ 0.4 has also been observed. In this article, we focus on high Sb-content ($x$~0.2) compounds if not specified.

The presence of magnetic rare earth elements provide a good opportunity to study the coupling between magnetism and topological quantum states [37,38]. AFM ground state and metamagnetic spin-flop transitions has been reported in CeSbTe [37,39] and GdSbTe [41]. For NdSbTe, we have also observed an AFM ground state. As shown in Fig. 2a, with the out-of-plane ($H\|c$) and in-plane ($H\|ab$) magnetic fields of 0.1 T, the temperature-dependent molar magnetic susceptibility $\chi_{mol}$ (= $M/H$) of NdSbTe displays a peak around 2.7 K, without any irreversibility (Fig. 2a, inset) between zero-field-cooling (ZFC) and field-cooling (FC). Among multiple samples we have measured, the transition temperature of 2.7 K appears insensitive to the composition stoichiometry. For $T > 50$ K, $\chi_{mol}$ in the paramagnetic state can be described by Curie-Weiss law $\chi_{mol} = \chi_0 + \frac{C}{T - \Theta}$ where $\chi_0$ is the temperature-independent part of susceptibility, $C$ is the Curie constant and $\Theta$ is the Weiss temperature, as shown in Fig. 2a. An effective magnetic moment of $\mu_{eff} = \sqrt{3k_B C / N_A} = 3.6\,\mu_B$ can be obtained from the fitted Curie constant, in agreement with the theoretical value of 3.62 $\mu_B$ for Nd$^{3+}$ with $4f^3$ configuration. The negative Weiss temperature of -17.8 K is comparable with that of CeSbTe (-10 ~ -23.9 K) [37,39] and GdSbTe (-19 ~ -24 K) [38,40,41], implying AFM exchange interaction between Nd moments.

The linear field dependence of the isothermal magnetization (Fig. 2b) in the low field region ($|\mu_0 H| < 1.4$ T) is also in line with an AFM ground state. At higher fields, deviation from linearity have been observed in $M(H)$ below the AFM ordering temperature $T_N = 2.7$ K. $M$ exhibits a steeper increase with field beyond $H_{c1} \approx 1.4$ T and evolves toward sublinear field dependence at higher field above $H_{c2} \approx 3.8$ T. These critical fields $H_{c1}$ and $H_{c2}$ do not show anisotropy and appear the same for $H\|c$ and $H\|ab$, though the $c$-axis should be the easy axis given the larger magnetization for $H\|c$. The observed magnetization behavior is most likely attributed to an AFM-to-canted AFM transition at $H_{c1}$ and a subsequent field-driven FM polarization at $H_{c2}$, which is reminiscent of that of the AFM topological insulator MnBi$_2$Te$_4$ [44]. Metamagnetic transitions have been probed in CeSbTe [37,39] and GdSbTe [41] which also exhibit AFM ground states. A fully polarized FM phase in CeSbTe can be reached at low field (< 1 T for $H\|c$) [37,39], while in GdSbTe field-driven moment canting has been suggested [41].

The metamagnetic transitions and FM polarizations provide a new approach to turn on/off the time-reversal symmetry, which has been proposed to modify the topological phases in CeSbTe [37].

The effect of an external magnetic field on the magnetic order in NdSbTe has also been probed with magnetic susceptibility measurements. As shown in the inset of Fig. 2a, upon increasing magnetic field, $T_N$ (indicated by dark triangles) shifts to lower temperatures, becoming unobservable down to 1.8 K (the lowest temperature can be reached by our PPMS) at $\mu_0 H = 5$ T. Such AFM order suppression is also seen in our specific heat measurements. In Fig. 3a we present the temperature dependence of specific heat $C(T)$ for NdSbTe. A sharp peak at $T_N = 2.7$ K at zero field can be observed, which is gradually suppressed and shifted to lower temperature upon the application of the magnetic field, disappearing for $\mu_0 H > 5$ T (Fig. 3a, inset).

Specific heat measurements also provide important details about the magnetic order. As shown in Fig. 3a, an external magnetic field modifies specific heat significantly even above $T_N$, leading to specific heat enhancement up to ~ 22 K, implying remarkable magnetic fluctuations ~20 K above the ordering temperature. The measured total specific heat $C_{tot}$ can be expressed as $C_{tot} = C_m + C_{el} + C_{ph}$, where $C_{el}$ and $C_{ph}$ represent the conventional electronic and phonon specific heat, respectively. Generally, separating each term is not difficult since $C_{el} = \gamma T$ and $C_{ph} = \beta T^3$ when $T \ll$ Debye temperature $\Theta_D$. However, in NdSbTe the magnetic fluctuations extend to high temperatures, as which the $T^3$-dependence for $C_{ph}$ become inaccurate. Alternatively, we adopt the structurally similar, non-magnetic LaSbTe as a reference sample to approach $C_{ph}$ for NdSbTe. As shown in Fig. 3b, the specific heat for LaSbTe (blue line) does not display any anomaly and follows conventional $C(T) = \gamma^{La} T + \beta^{La} T^3$ with $\gamma^{La} = 0.51$ mJ/mol K$^2$ and $\beta^{La} = 0.40$ mJ/mol K$^4$ from 1.8 to 3 K (Fig. 3b, inset), so the phonon specific heat of LaSbTe in the full temperature range can be extracted by $C_{ph}^{La}(T) = C(T) - \gamma^{La}T$. According to the corresponding state principle [45], the phonon entropy $S_{ph}$ for NdSbTe and LaSbTe can be expressed by a universal function $f(T/\theta)$ where $\theta$ is the material-dependent parameter. Therefore, the phonon specific heat of NdSbTe can be expressed by that of LaSbTe by $C_{ph}^{Nd}(T) = A \cdot C_{ph}^{La}(B \cdot T)$, where A and B are renormalization factors. The specific heat for NdSbTe can thus be represented by $C^{Nd}(T) = C_m^{Nd}(T) + C_{el}^{Nd}(T) + C_{ph}^{Nd}(T) = C_m^{Nd}(T) + \gamma^{Nd}T + A \cdot C_{ph}^{La}(B \cdot T)$, which reduces to $C^{Nd}(T) = \gamma^{Nd}T + A \cdot C_{ph}^{La}(B \cdot T)$ in the high temperature paramagnetic phase where $C_m$ is negligible.

With the known $C_{ph}^{La}(T)$ we fit the high temperatures specific heat for NdSbTe (Fig. 3b, black line). The fitting parameter $\gamma^{Nd}$, the Sommerfeld coefficient, is 115 mJ/mol K$^2$ (Fig. 3b, inset) which is much greater than other known $Ln$SbTe compounds such as LaSbTe (0.51 mJ/mol K$^2$, see inset of Fig. 3b), GdSbTe (7.6 mJ/mol K$^2$) [41], and CeSbTe (10-40 mJ/mol K$^2$) [37,39], implying greatly enhanced electron correlation in this material. The fitted renormalization factors of A = 1.06 and B = 0.98 for phonon specific heat are reasonably close to 1 given the structure similarity between NdSbTe and reference compound LaSbTe. This leads to $\beta^{Nd}$ = 0.44 mJ/mol K$^4$ and a Debye temperature of $\Theta_D = (12\pi^4 NR/5\beta)^{1/3}$ = 236.6 K with atom number per formula cell $N$ = 3 and gas constant $R$ = 8.31 J/mol K, which is slightly lower than that of 244.3K for LaSbTe ($\beta^{La}$ = 0.40 mJ/mol K$^4$).

Magnetic specific heat due to the Nd moment can be extracted by subtracting the fitted paramagnetic background from the total measured data, as depicted in Fig. 3c. Above $T_N$, a long tail extending to ~20 K can be seen, which is attributed to the magnetic fluctuations. The magnetic entropy $S_m = \int_0^T \frac{C_m(T)}{T} dT$ saturates to 9.1 J/mol K at 20K, which is about 79% (2.4

J/mol K less) of the expected value of $R\ln(2S+1)$ with $S = 3/2$ for $Nd^{3+}$. Deviation from the expected $S_m$ is also seen for GdSbTe [41] but not in CeSbTe [37,39]. Further studies including millikelvin specific heat measurements and theoretical efforts are needed to clarify such large entropy reduction.

The coupling between the magnetism and transport properties have been studied in this work. As shown in Fig. 4a, the overall temperature dependence of resistivity for NdSbTe exhibits non-metallic behavior with increasing resistivity upon cooling, with a small peak centered at ~3.2 K which can be suppressed by magnetic field (Fig. 4a, inset). Given the proximity of the peak temperature and the AFM ordering temperature ($T_N$ = 2.7 K), as well as the similar field suppression of AFM order observed in magnetic susceptibility and specific heat, the suppression of the resistivity peak is most likely due to the suppression of spin scattering. It is also worth noting that NdSbTe appears to slowly degrade over time, and this low temperature peak disappears after a few weeks. At higher temperatures, resistivity displays a logarithmic temperature dependence up to 50 K, as shown in Fig. 4a (indicated by dashed lines). The logarithmic temperature dependence has also been observed in CeSbTe in a much narrower temperature range (3 ~ 7 K), which has been ascribed to Kondo effect [39].

Although a multiband nature has been observed for the known $Ln$SbTe compounds [37,38], NdSbTe displays linear field dependence for Hall resistivity $\rho_{xy}(H)$ in the paramagnetic state with negative slope (Fig. 4b, inset), indicating that the transport is dominated by the electron band. Single-band behavior has also been found in CeSbTe [39] despite of its multiband nature [37], though it is very different from that of the non-magnetic ZrSiS-type compounds in which the multiband nature is clearly manifested by non-linear $\rho_{xy}(H)$ even at relatively high temperatures [27,36,46-48]. The Hall coefficient extracted by $R_H = d\rho_{xy}/d(\mu_0 H)$ is depicted in Fig.

4b, which reaches a minimum at 20 K. The corresponding electron density is estimated to be on the order of $1\times10^{21}$ cm$^{-3}$ (Fig. 4b). This carrier density is much lower than that of conventional metals, but is comparable with CeSbTe [39] and other non-magnetic ZrSiS-type semimetals [27,31,47]. From the obtained Hall coefficient and the zero-field resistivity, the mobility estimated from a single-band model $\mu = R_H/\rho$ is found to be as low as 2 ~ 3 cm$^2$/Vs, in agreement with the non-metallic transport behavior.

In Fig. 4c we show the magnetoresistance (MR) at different temperatures. The field dependence of resistivity at $T$ = 2 K exhibits a sharp drop and reaches minimum around $\mu_0H$ = 1.1 T, which is consistent with the field-suppression for low temperature resistivity peak (Fig. 4a, inset.) The field dependence of the resistivity is weak between 1.1 T and 3.7 T, and evolves to negative MR at higher fields. Here the critical fields of 1.1 T and 3.7 T correspond well with the metamagnetic transition of the magnetization ($H_{c1}$ = 1.4 T and $H_{c2}$ = 3.8 T, Fig. 2b), suggesting that the MR behaviors at 2 K may be associated with spin scattering. Above $T_N$, the MR peak at zero field disappears (Fig. 4c), and MR displays a crossover from positive to negative with increasing field. Above 20 K, MR is negative in the full field range up to 9 T. It is worth noting that around 20 K the Hall coefficient also reaches a minimum (Fig. 4b), and the magnetic fluctuations start to develop (Fig. 2a).

The negative MR in NdSbTe is found to be weak (0.8% at 9 T and 2 K) and not strongly dependent on the magnetic field orientation. As shown in Fig. 4d, when the magnetic field rotates from the out-of-plane ($\theta$ = 0°, see inset of Fig. 4d) to the in-plane ($\theta$ = 90°) directions, the negative MR in the high field region only becomes slightly larger, possibly due to the suppression of the weak positive orbital MR component as $H$ is rotated toward the current direction. The negative longitudinal MR when $H||I$ ($\theta$ = 90°) is a signature for Weyl states as it is associated with the chiral magnetic effect of Weyl fermions. However, the MR in NdSbTe is negative for arbitrary field orientations without remarkable evolution with $\theta$, which can be ascribed to the field suppression of spin scattering. Therefore, the chiral anomaly-induced negative longitudinal MR, if existent, is not apparent.

We now discuss the interplay between magnetism and the electronic bands in NdSbTe. As a special group of the ZrSiS-type topological material family, the magnetism from the *Ln* in *Ln*SbTe is expected to couple with the electronic band topology and induce various topological electronic states. This has been supported by ARPES observations and band structure calculations in CeSbTe [37] and GdSbTe [38]. Observation of the topological phase transition under magnetic field is not experimentally feasible with ARPES, and magnetotransport measurements would be the most effective approach. However, transport evidence for non-trivial topology in *Ln*SbTe, including the chiral anomaly and quantum oscillations with non-trivial Berry phase, have yet to be discovered due to the lack of magnetotransport study. In NdSbTe, the similarity in structure and magnetism with CeSbTe and GdSbTe implies a similar interplay between magnetism and band topology. Particularly, the FM polarization beyond 3.8 T and the sizable magnetic moment of NdSbTe (Fig. 2b) suggest that the time-reversal symmetry breaking Weyl fermions may arise in a similar manner as that predicted for CeSbTe [37]. As stated above, our magnetotransport indeed reveal the interplay between magnetism and electronic transport properties, but chiral anomaly-induced negative MR is not sufficiently conclusive. In addition, the observation of quantum oscillation in NdSbTe is difficult due to its low mobility. Given the trend of field-driven FM state seen in magnetization measurements, it would be helpful to push to high field to search for possible quantum oscillation and signatures of topological fermions in this material. The full moment polarization at high field would be favorable for a time-reversal symmetry breaking Weyl state, which has been demonstrated in Mn(Bi,Sb)$_2$Te$_4$ system [49-51].

So far, the reported magnetic *Ln*SbTe compounds including CeSbTe and GdSbTe [39,41] exhibit non-metallic transport, despite their metal-like electronic band structures [37,38]. Charge-density-wave (CDW) in orthorhombically distorted GdSbTe [40] has been reported. The orthorhombic distortion and the associated CDW are found to be associated with Sb-deficiency in GdSb$_{1-x}$Te$_{1+x}$ and disappear for $x \leqslant 0.15$ [40], but the nearly stoichiometric GdSb$_{0.997}$Te$_{1.003}$ still exhibits non-metallic transport behavior [41]. Therefore, further structure and transport studies on samples with different Sb content are needed to clarify a CDW scenario in NdSbTe. In addition to CDW, A Kondo mechanism has also been claimed for CeSbTe [39]. In NdSbTe, the logarithmic temperature dependence of resistivity and weak angular dependence of MR are suggestive of a Kondo origin. Though typical Kondo localization in dilute magnetic alloys usually occurs at much lower temperatures, non-metallic transport extending to high temperatures can be found in Kondo insulators such as Ce$_3$Bi$_4$Pt$_3$ [52]. Indeed, the strong electronic correlations evidenced by the large Sommerfeld coefficient (115 mJ/mol K$^2$, largest among the known *Ln*SbTe compounds) may be ascribed to the mass enhancement due to Kondo hybridization of the *f*-electrons and the conduction electrons. Nevertheless, the band structure calculations for CeSbTe indicates that the 4*f*-band is far away from the Fermi level [37], so the scenario of hybridization in NdSbTe needs further theoretical investigations. In addition, ZrSiS-type compounds are known as topological nodal-line semimetals, in which the reduced kinetic energy along the line may give rise to strong correlations. Enhanced correlations have been probed in a few non-magnetic ZrSiS-type materials including ZrSiS [53] and ZrSiSe [54] despite the small Sommerfeld coefficients [55]. As a special group of ZrSiS-family, *Ln*SbTe compounds can also support Dirac nodal-line crossings as has been discovered in GdSbTe [38]. Therefore, the coupling between nodal-line fermions, Kondo hybridization, and magnetism could generate novel correlated phenomena in NdSbTe.

In summary, we have successfully synthesized single crystals of NdSbTe, a new magnetic ZrSiS-type compound, and characterized the magnetic, electronic transport, and calorimetric properties. This material exhibits interesting coexistence of AFM order with metamagnetic transitions, and possible Kondo localization. Although the topological bands in NdSbTe remain elusive, this material provide an ideal platform for investigating rich physical phenomena including the interplay between magnetism and topological fermions, the nonsymmorphic Kondo-Weyl state [56], and correlated electron physics in topological quantum materials.

**Acknowledgement**

This work is primarily supported by the U.S. Department of Energy, Office of Science, Basic Energy Sciences program under Award No. DE-SC0019467. Part of the material characterizations (XRD and EDS) is supported by Arkansas Biosciences Institute. The authors thank Prof. S. Barraza-Lopez and Dr. J. Villanova from the University of Arkansas for informative discussions.

**Figures**

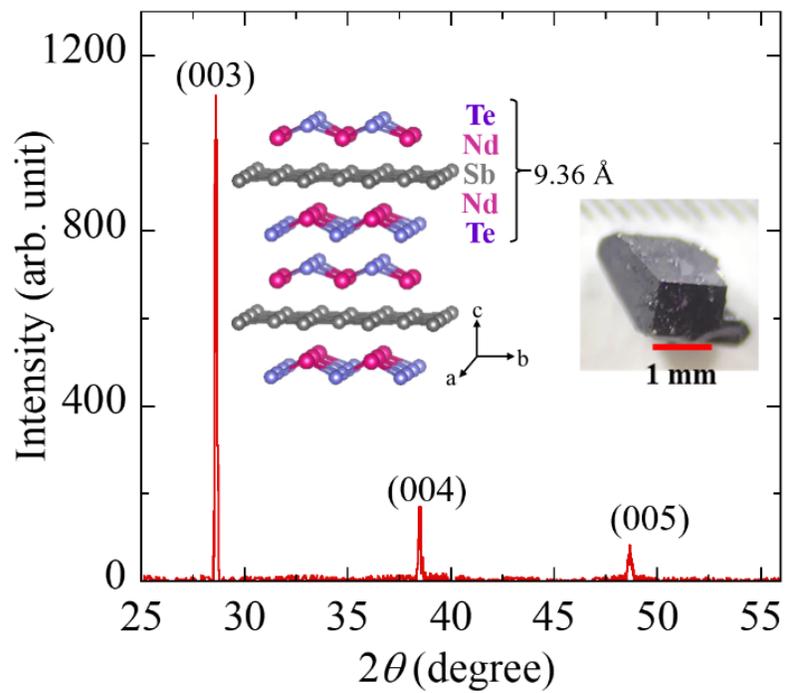

**Figure 1.** Single crystal x-ray diffraction pattern for NdSbTe, showing the (00*L*) reflections. Insets: crystal structure and an optical microscope image of a NdSbTe single crystal.

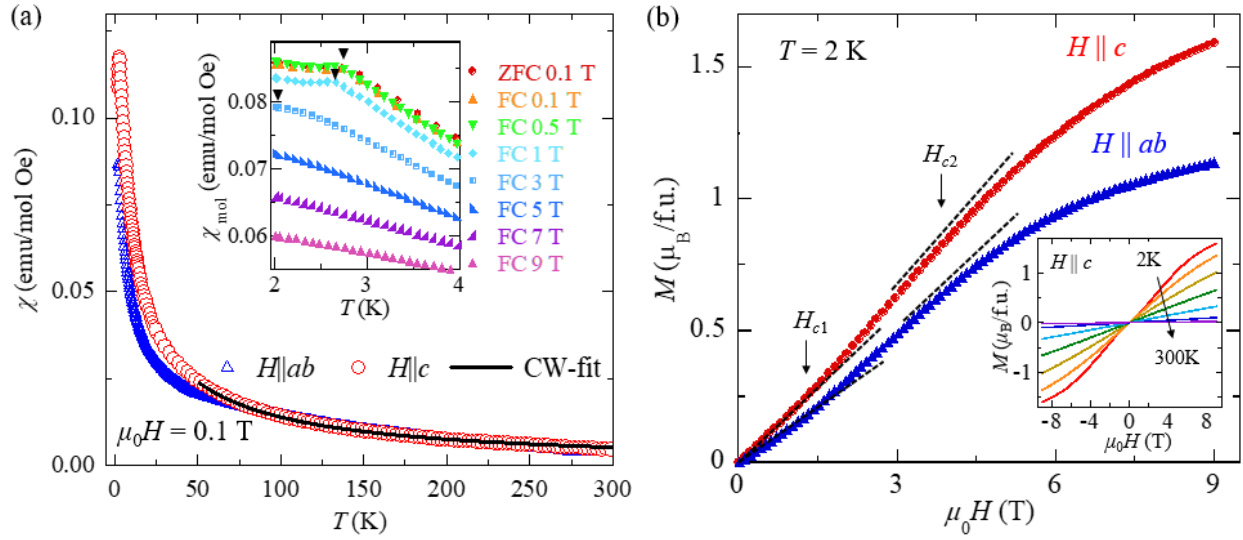

**Figure 2.** Magnetic properties of NdSbTe. (a) Temperature dependence of molar susceptibility of NdSbTe measured under out-of-plane (*H*ll*c*) and in-plane (*H*ll*ab*) magnetic fields of 0.1 T. The solid line represents the Curie-Weiss (CW) fit. Inset: magnetic susceptibility taken under different fields from 0.1 to 9 T. The absence of irreversibility between ZFC and FC at 0.1 T is shown. (b) Field dependence of magnetization at 2 K for *H*ll*c* and *H*ll*ab*. The dashed lines are guides for eyes. Insets: magnetization at $T = 2, 5, 10, 20, 50, 150$, and 300 K for *H*ll*c*.

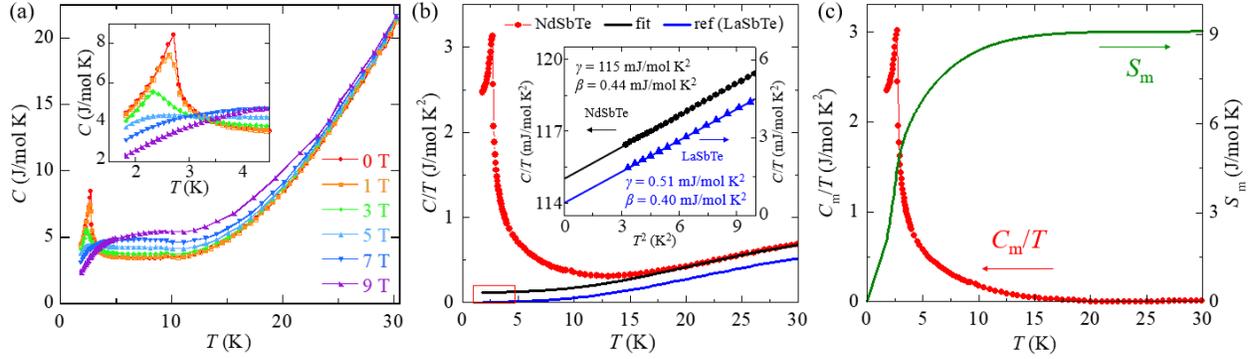

**Figure 3.** Specific heat of NdSbTe. (a) Temperature dependence of specific heat $C$, measured under different magnetic from 0 to 9 T. The magnetic field is applied along the $c$-axis. Inset: zoom-in of the low temperature specific heat. (b) Specific heat divided by temperature $C/T$ for NdSbTe (red) and the reference sample LaSbTe (blue). The black curve is the fit according to the corresponding principle (see text). The zoom-in of low temperature data in the red box is shown in the inset, but plotted in $C/T$ vs. $T^2$. The solid line represents the linear fits to $C/T = \gamma+\beta T^2$. (c) Temperature dependences of magnetic specific heat divided by temperature $C_m/T$ and the magnetic entropy $S_m$.

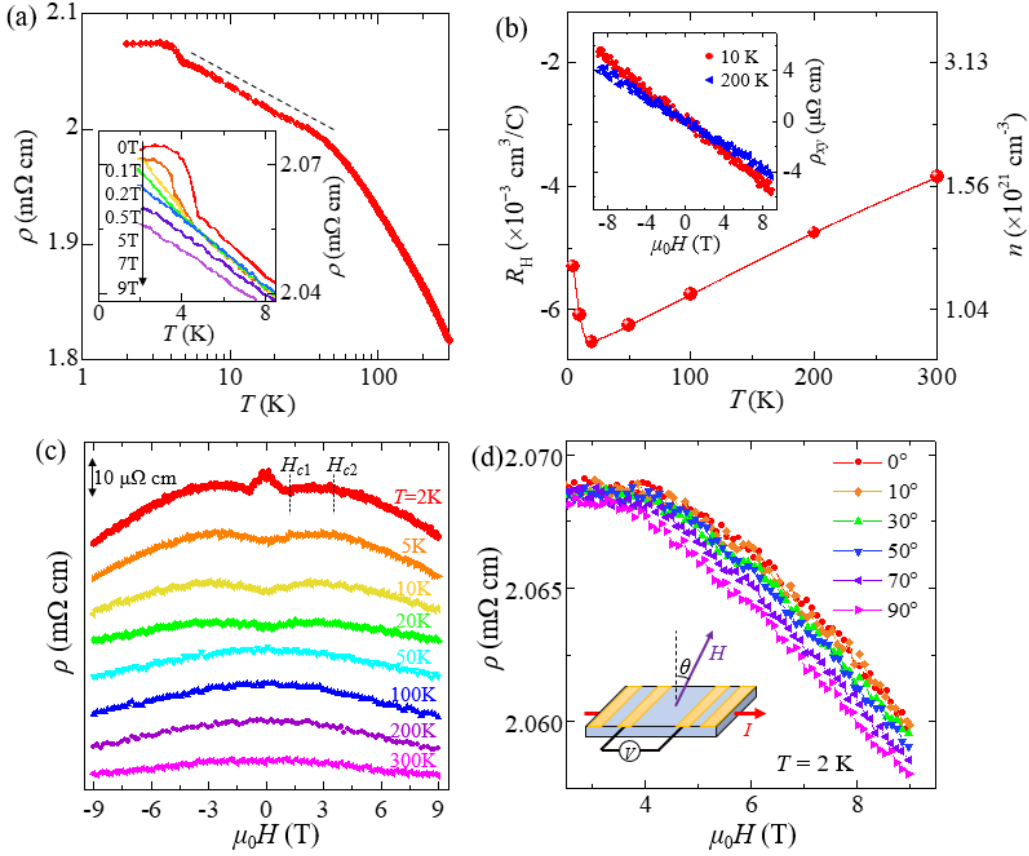

**Figure 4.** Electronic transport properties of NdSbTe. (a) Temperature dependence of in-plane resistivity of NdSbTe. Logarithmic temperature dependence can be observed 5 − 50 K. The dashed lines are guides for eyes. Inset: low temperature resistivity under different magnetic fields. (b) Temperature dependence of Hall coefficient $R_H$ and the corresponding carrier density ($1/eR_H$). Inset: field dependence of Hall resistivity at 10 and 100 K. (c) Field dependence of in-plane resistivity at different temperatures. Data for different temperatures are shifted for clarity. $H_{c1}$ and $H_{c2}$ are critical fields determined from magnetization (Fig. 2b). (d) Field dependence of resistivity at 2 K under different magnetic field orientations. High field data (above 2.5 T) is shown for clarity. Inset: the measurement setup.